\documentclass[aps,twocolumn,nofootinbib,preprintnumbers,superscriptaddress,eqsecnum]{revtex4}
\usepackage[
      colorlinks=true,
      linkcolor=blue,
      urlcolor=blue,
      filecolor=black,
      citecolor=red,
      pdfstartview=FitV,
      pdftitle={},
        pdfauthor={Veronika Hubeny},
        pdfsubject={},
        pdfkeywords={},
        pdfpagemode=None,
        bookmarksopen=true
      ]{hyperref}

\usepackage[normalem]{ulem}
\usepackage{amsmath}
\usepackage{enumerate}
\usepackage{amsfonts}
\usepackage{color}
\usepackage{graphicx}
\usepackage{bm}

\def\transv{{\vec V}_{\perp}}
\def\CN{{\cal N}}
\def\CE{{\cal E}}

\begin{document}

\title{Relativistic Beaming in AdS/CFT}
\preprint{DCPT-10/53}
\author{Veronika E. Hubeny}
\affiliation{Centre for Particle Theory \& Department of Mathematical Sciences, Durham University,
Science Laboratories, South Road, Durham DH1 3LE, United Kingdom}

\begin{abstract}

We propose a mechanism of `beaming' the backreaction of a relativistic source in the bulk of AdS towards the boundary.
Using this beaming mechanism to estimate the energy distribution from radiation by a 
circling quark 
in strongly coupled field theory, we find remarkable agreement with the previous results of \cite{Athanasiou:2010pv}. 
Apart from explaining a puzzling feature of these results and elucidating the scale/radius duality in AdS/CFT, our proposal provides a useful computational technique.

\end{abstract}

\today

\maketitle

\section{Introduction}
\label{s:intro}

The AdS/CFT duality has been paramount in elucidating strongly coupled gauge dynamics on the one hand, and quantum gravity on the other.
One useful strategy has been to recast the hard questions from one side of the correspondence into the dual language 
which typically circumvents insurmountable obstacles encountered by a direct approach.
However, efficient implementation of such a program relies on good understanding of the dictionary between the two sides.  

One crucial entry in the AdS/CFT dictionary is the scale/radius (or UV/IR) duality \cite{Susskind:1998dq}, suggesting that the closer a bulk excitation is to the AdS boundary, the smaller is the characteristic size of the dual CFT excitation.  This provides an important insight into the holographic nature of the duality: the `extra' (radial) dimension 
in the bulk arises from a scale in the dual CFT.  This intuition 
bolsters our understanding of 
various dynamical processes.  For example, the field theory notion of color transparency, that different-scale excitations pass through each other without interacting, is beautifully explained by bulk locality: in the bulk, the corresponding excitations do not interact due to being separated in the radial direction.

However, naive applications of the scale/radius duality should be treated with caution.  
In particular, it is {\it not}  always true that different-scale excitations on the boundary do not interact.  Nor is the converse true, that same-scale excitations passing through each other in the CFT do interact.  At the heart of both these observations lies the realization that {\it a source deep in the bulk of AdS may nevertheless produce a sharply-localized effect on the boundary.}  
One recent example of such a phenomenon appeared in the context of \cite{Athanasiou:2010pv} (summarized by \cite{Athanasiou:2010ss}), where a coiling string in AdS gives rise to a synchrotron-radiation-like effect on the boundary.  The results of \cite{Athanasiou:2010pv} suggest that the backreaction of the string remains sharply localized in the bulk geometry all the way to the boundary, as if it were `beamed'.

In this paper, 
we propose a mechanism for such a `beaming' effect, and test it in the context of  \cite{Athanasiou:2010pv}.  
This not only elucidates the nature of the scale/radius duality at a deeper level, but as we explain below, it also provides a convenient and simple method to calculate the radiation from a relativistic quark following an arbitrarily accelerating trajectory in  a strongly-coupled medium. 
Below we summarize the important results, leaving further details to a longer companion paper \cite{Hubeny:2010xx}.

\section{Rotating helical string and synchrotron radiation}
\label{s:setup}

The authors of \cite{Athanasiou:2010pv} use the AdS/CFT correspondence to study the radiation from a quark in uniform circular motion in a strongly coupled field theory.\footnote{
They achieve this by utilizing the well-known feature of AdS/CFT, that classical gravity in the bulk captures the full quantum effects of the field theory. 
A test quark is modeled by an infinitely massive spin $\frac{1}{2}$ particle of an $\CN = 2$ hypermultiplet.
It is coupled to $N_c \to \infty$ limit of SYM, at finite 't Hooft coupling $\lambda \gg 1$.
To obtain explicit expressions  which can be directly compared to the behavior at weak coupling, 
\cite{Athanasiou:2010pv} consider the SYM in its vacuum state.
} 
 In the gravitational dual, such set-up corresponds to a string in AdS, 
 ending on the quark.  
The gravitational backreaction of the string deforms the bulk geometry away from AdS, and this in turn induces a non-trivial stress tensor on the boundary, which captures both the motion of the quark, as well as the resulting radiation.  

To state the results more explicitly, let us first set the notation.
Consider a test quark in the boundary field theory moving on a circle of radius $R_0$ with constant angular velocity $\omega_0$.
The linear velocity of the quark is then $v=R_0 \, \omega_0$, with the corresponding Lorentz factor $\gamma \equiv \frac{1}{\sqrt{1-v^2}}$.
The gravitational dual corresponds to a trailing string moving in AdS.  In the Poincare coordinates  $(t,r,\theta,\varphi,u)$, where  AdS (of unit size) has the metric
\begin{equation}
ds^2 = \frac{1}{u^2} 
\left[ - dt^2 + dr^2 + r^2  \left( d\theta^2 + \sin^2 \theta \, d \varphi^2 \right) + du^2 \right] \ ,
\label{e:AdSmet}
\end{equation}	
$u$ corresponds to the bulk radial coordinate ($u=0$ is the AdS boundary) and $r$ denotes the boundary-radial coordinate.
Parameterizing the string worldsheet by $t$ and  $u$, \cite{Athanasiou:2010pv} write the string embedding as 
\begin{equation}
X^{M}(t,u) = (t,R(u), \frac{\pi}{2},\phi(u) +\omega_0 \, t , u) \ , 
\qquad
\label{e:Xcoords}
\end{equation}	
and solve the Nambu-Goto action to determine the profile functions $R(u)$ and $\phi(u)$: 
\begin{eqnarray}
R(u) &=& \sqrt{R_0^2 + v^2 \, \gamma^2 \,  u^2} \\
\phi(u) &=& - u  \, \gamma  \, \omega_0 + \tan^{-1} \left( u  \, \gamma  \, \omega_0 \right) \ .
\label{string}
\end{eqnarray}	
The string then has the shape of a  rigidly rotating helix, flaring-out into the bulk at a rate $R'(u) \approx v  \, \gamma$.

To find how the radiation induced by the quark's acceleration propagates through the strongly coupled medium, 
\cite{Athanasiou:2010pv} consider the energy density $\CE \equiv T^{tt}$, where $T^{\mu\nu}$ is the boundary stress tensor. 
Since the on-shell bulk gravitational action is the generating functional for the boundary stress tensor, $T^{\mu\nu}$ can be read-off from the asymptotic form of the bulk metric deformation.  In particular, using the bulk stress tensor produced by the string, \cite{Athanasiou:2010pv} solve the linearized Einstein's equations near the boundary.
This yields an analytic few-line expression for $\CE(t,r,\theta,\varphi)$  (see eq.(3.70) of \cite{Athanasiou:2010pv}, plotted in their Fig.4) in terms of retarded time $t_{\rm ret}$ which is evaluated numerically. 

This boundary energy density turns out to have rather surprising properties:  it is mainly supported on a thin spiral  rotating rigidly at constant angular velocity, with the energy pulses {\it not} widening in the boundary-radial direction.\footnote{
This statement holds only at $T=0$; at non-zero temperature, these pulses would widen, slow down, and eventually thermalize.  However, for $\omega_0^2 \, \gamma^3 \gg \pi^2 \, T^2$, the synchrotron radiation persists on length scales smaller than the thermal scale $1/T$.} 
Indeed, the radiation pattern is in quantitative agreement with corresponding radiation at weak coupling, which mimics that of synchrotron radiation in classical electrodynamics.

More specifically, the energy density $\CE$ exhibits the following salient features.
\begin{itemize}
\item Each pulse remains constant width as it propagates outward, i.e.\ it does not broaden.
\item A pulse propagates outward at the speed of light, independently of the quark velocity $v$.  (This means that the spatial separation between the spiral arms of $\CE$ at a fixed time is $\sim 2 \pi \, R_0 / v$.)
\item Pulse width decreases with increasing $v$ as $\sim 1/\gamma^3$.
\item $\CE$ decreases off the orbital plane, with the characteristic width narrowing for increasing $v$ as $\sim 1/\gamma$.
\end{itemize}
As argued in \cite{Athanasiou:2010pv}, these features -- especially the first point -- are quite surprising from the field theory point of view: since the radiation is strongly coupled, one might have expected it to diffuse and isotropize.

\section{Proposal for beaming mechanism}
\label{s:beaming}

The curious features 
summarized above are no less surprising and fascinating from the bulk standpoint: why does the gravitational backreaction of the bulk string behave in such a sharply-localized fashion?  As pointed out in the Introduction, this is at odds with the naive expectations from the scale/radius duality that the deeper parts of the string would produce more diffuse signals on the boundary.  In this section, we propose a general mechanism which leads to the observed localization; we refer to this mechanism as `beaming'. 

We start by noting that large part of the string is moving relativistically.
 Since the string has no longitudinal excitations, a natural quantity to describe the string's motion is the transverse velocity of the string, $\transv(t,u)$.  
The direction of the transverse velocity\footnote{
Although the exact expression is lengthy, at large $\gamma$ and $u/R_0$, the leading $(r,\phi,u)$ components of $\transv{}$ approach $\left( 1 ,\frac{R_0}{\gamma^2 \, u^2} ,\frac{1}{\gamma} \right)$.
}
 points towards larger $r$, $\phi$, and $u$, i.e.\ away from the AdS boundary,
with its magnitude given by
\begin{equation}
||\transv(u)|| =  \sqrt{\frac{ v^2 \, R_0^2 + v^4 \, \gamma^4 \, u^2}{R_0^2 + v^4 \, \gamma^4 \, u^2}} \ ,
\label{e:transvel}
\end{equation}	
which 
increases monotonically with $u$, and asymptotes  1 deep in the bulk.

The object then is to understand how such a relativistic string backreacts on the spacetime.  We use the known fact that the backreaction of a massless particle moving at the speed of light is given by a {\it gravitational shock wave} (GSW) \cite{Aichelburg:1970dh}.
Such metric perturbation is supported\footnote{
The actual construction of a GSW in AdS can be obtained similarly to \cite{Aichelburg:1970dh}, by taking a double-scaling limit of a boosted black hole, with the boost parameter taken to infinity and mass to zero, keeping the total energy fixed \cite{Hotta:1992qy}.} on a null plane transverse to the particle's velocity, analogously to the behaviour of the electric field for an infinitely-boosted charge in classical electrodynamics.
Accordingly, the backreaction of all `string bits' (viewed as a continuum of particles distributed along the string), moving relativistically in the direction of $\transv$, should be given by a superposition of such GSWs, one for each particle.
We therefore propose that {\it the backreaction the full string is well-approximated by a linear superposition of gravitational shock waves normal to the string's transverse velocity.}

This proposal rests on several important assumptions.  Firstly, we require that the interaction between the individual string bits does not contribute appreciably to the backreaction of the string.  Admittedly such an assumption would fail for a slowly-moving string where longitudinal boost invariance of the string's stress tensor implies that the (negative) pressure is of the same magnitude as the energy density.  However, under transverse boosts with Lorentz factor  $\gamma_{\perp}$, energy density is enhanced  by a factor of $\gamma_{\perp}^2$ compared to the pressure which captures the effect of interactions.  In fact, this is not surprising, given the parton model of the string \cite{Kogut:1972di}. 
This means that our assumption of non-interaction becomes arbitrarily good for high enough string's transverse velocity, so that in the relativistic regime we can indeed treat the string as composed of relativistic particles.  

The second assumption is that each string bit produces a backreaction in the form of GSW, which is valid as long as the string's transverse velocity is sufficiently relativistic.  Since from eq.(\ref{e:transvel}), we find that 
$V_{\perp}^2 \, \gamma_{\perp}^2 = \gamma^2 \, (v^2+u^2/u_h^2)$, 
with $u_h \equiv \frac{R_0}{v^2 \, \gamma^2}$ denoting the worldsheet horizon,
our second assumption holds sufficiently deep in the bulk or everywhere for sufficiently relativistic quark.

Finally, we assume that the individual GSWs superpose linearly, despite the nonlinearities of general relativity.  Although this may not hold for arbitrary GSWs -- for example head-on collisions can produce black holes \cite{Grumiller:2008va} -- it does hold in our case: examining the geometry in detail, we can show that the individual GSWs intersect mildly enough (in the sense indicated in the next section), and their interaction is strongest at the string itself.  Because we started with a nearly-test string, the GSWs produced by its backreaction will themselves backreact parametrically less (by ${\mathcal O}(1/N_c) \to 0$ in the present context).
This justifies superposing individual GSWs to estimate the backreaction of the string.

\section{Tests of beaming mechanism}
\label{s:tests}

In the previous section we have proposed that the backreaction of the string is given by a linear superposition of GSWs normal to the string's transverse velocity, and we have indicated justifications of the assumptions underlying this mechanism.  Nevertheless, justifications aside, the real test is whether the mechanism actually works: does such a superposition of GSWs lead to the energy density $\CE$ calculated by \cite{Athanasiou:2010pv}?
This section summarizes the critical tests we have carried out (with further details relegated to \cite{Hubeny:2010xx}).  We will see that, indeed, the boundary energy density arising from the GSW superposition is in remarkable quantitative agreement with all of the salient features of $\CE$ found by \cite{Athanasiou:2010pv} (as summarized in Section \ref{s:setup}).

Let us first consider the GSW corresponding to a single string bit, parameterized by time $t$ and bulk radius $u$.  It must be localized on the transverse `plane', which in coordinates of (\ref{e:AdSmet}) has the shape of a hemisphere with its equator lying on the boundary.  We can construct such a shock wave by emanating spacelike constant-$t$ geodesics from the given point $(t,u)$ on the string normally to $\transv(t,u)$.  
The boundary `light-up' from each GSW is then supported on a sphere of radius
$\rho$ and origin  at $(r_o, \theta_o = \frac{\pi}{2}, \varphi_o)$, 
 given approximately\footnote{
The expressions presented here pertain to the regime $u \gg R_0$; the exact full expressions have been calculated but are too lengthy and unilluminating to present here.
} 
by 
\begin{equation}
\rho \approx \sqrt{\gamma^2 \, u^2 +\frac{R_0^2}{v^4}} \ , \ \ 
r_o \approx \frac{R_0}{v^2} \ , \ \ 
\varphi_o \approx \omega_0 ( t -\gamma \, u) \ .
\label{e:singleGSW}
\end{equation}	
Of course, the amplitude of the GSW will be strongest nearest to the source, and will fall-off rapidly away from this direction; within the boundary, this dominant light-up due to the string bit at $(t,u)$ occurs at 
\begin{equation}
r \approx \gamma \, u \ , \ \ 
\theta = \frac{\pi}{2} \ , \ \ 
{\rm and} \ \
\varphi \approx \omega_0 ( t - \gamma \, u)+ \frac{\pi}{2}\ .
\label{e:GSWlightup}
\end{equation}	

Having understood the geometry of each individual GSW, we can now consider their superposition.
Clearly, there will be an enhancement where the GSWs are most dense, in particular where they intersect.
In the present geometrical set-up, this direction turns out to coincide with that characterizing the dominant effect of each GSW, (\ref{e:GSWlightup}).
\begin{figure}
\begin{center}
\includegraphics[width=3.2in]{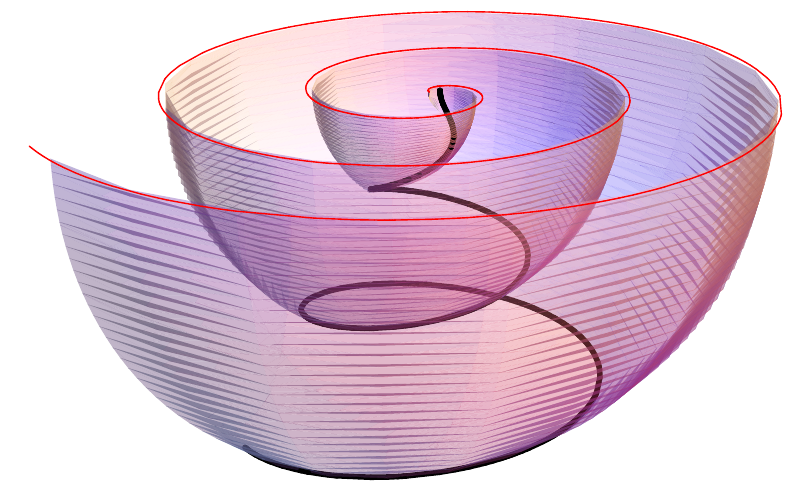}
\caption{Bulk profile of  GSW superposition for $v=0.5$, whose intersection with the boundary gives approximately the spiral curve (\ref{e:GSWlightup}) (thin red curve at the top).  The thick (black) curve denotes the string;  $r$ points horizontally, $u$ vertically down.
}
\label{f:bulk_GSW}
\end{center}
\end{figure}
Figure \ref{f:bulk_GSW} shows the bulk surface around which such an intersection is supported, plotted in the $(r,\varphi,u)$ space at $\theta = \pi/2$ and $t=0$.  For orientation we also include the string (thick black curve).  The boundary is at the top, and we can see that the boundary light-up due to the superposition of all GSWs will be clustered around a spiral, well-approximated by the relations (\ref{e:GSWlightup}) (thin red curve), which at fixed $t$ can be viewed as a curve on the orbital plane, parameterized by $u$.

From the expression (\ref{e:GSWlightup}) we immediately see that at any given time, the boundary light-up indeed has the expected spiral shape, with spacing between the spiral arms given by $\Delta r \approx 2 \, \pi/\omega_0 = 2 \, \pi \, R_0 / v$, reproducing the corresponding result in \cite{Athanasiou:2010pv}.  Moreover, since time periodicity (corresponding to $\Delta \varphi = 2 \pi$ in (\ref{e:GSWlightup})) is $\Delta t =  2 \, \pi/\omega_0$, this simultaneously implies that the energy pulse propagates outward at the speed of light, independently of the quark velocity $v$.  Furthermore, we can also confirm that the phase of the spiral agrees with that computed in \cite{Athanasiou:2010pv}, which corresponds to the `synchrotron' radiation being forward-beamed in the direction of the quark velocity.

While the above results demonstrate quantitative agreement with the second point listed in Section \ref{s:setup}, we wish to verify the other salient features of $\CE$ as well.  In order to find the width of the spiral arms, as well as the extent off the orbital plane, we need to consider the entire GSW (\ref{e:singleGSW}) rather than just its dominant contribution (\ref{e:GSWlightup}).  
\begin{figure}
\begin{center}
\includegraphics[width=2.8in]{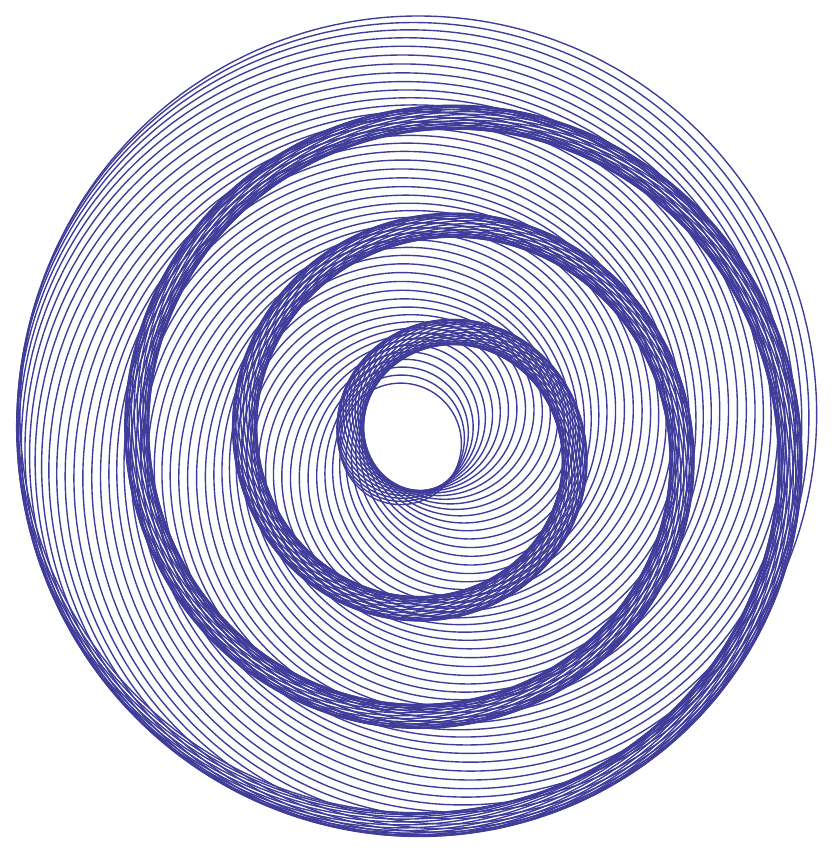}
\caption{Light-up from GSW originating at various points along the string (corresponding to quark velocity $v=0.5$ and $u\in(5,40)$ in increments of 0.3).  Each GSW forms a circle approximated by (\ref{e:singleGSW}); but from the superposition of all such circles, there clearly emerges an overall spiral pattern.}
\label{f:bdy_GSW}
\end{center}
\end{figure}
This is most easily analyzed if 
we focus on the support of the GSW and instead ignore the variations of its intensity.
Figure \ref{f:bdy_GSW} shows the resulting boundary light-up at $t=0$, $\theta = \pi/2$, and $u=0$, for quark velocity $v=0.5$.   While only the individual GSWs are plotted, each simply a circle with radius and origin given by (\ref{e:singleGSW}), we see that an overall spiral pattern emerges from the superposition of such circles.  Overlaying the spiral (\ref{e:GSWlightup}) plotted in Figure \ref{f:bulk_GSW}, we can verify that they match perfectly.

Most strikingly, Figure \ref{f:bdy_GSW} demonstrates that the spiral width (corresponding to the energy pulse discussed in Section \ref{s:setup}) does {\it not} broaden with radius $r$, but rather stays at constant width.  This width $\Delta$ of the spiral arm, which can be extracted from the curves corresponding to intersections of `adjoining' GSWs, is given by
\begin{equation}
\Delta = \frac{2 \, R_0}{v} \left( \frac{1}{v \, \gamma} - \tan^{-1}  \frac{1}{v \, \gamma} \right)
\ \ \
\xrightarrow[{v \to 1}]{}
\ \ \
 \frac{2 \, R_0}{3 \, v^4} \, \frac{1}{\gamma^3} \ .
 \label{}
\end{equation}	
This confirms the $\gamma^{-3}$ scaling indicated in third point in Section \ref{s:setup}.
A similar analysis of the GSW superposition off the orbital plane confirms the fourth point, namely that $\CE$ falls off with the polar angle $\theta$, with characteristic spread decreasing with increasing $v$ as $\Delta \theta \sim 1 / \gamma$.

Finally, we can combine the calculations summarized in Figures \ref{f:bulk_GSW} and \ref{f:bdy_GSW}, and consider the superposition of full GSWs, taking into account the falloff of the intensity, which near the dominant point (\ref{e:singleGSW}) decreases as
$(u^2 + \delta\varphi^2 + \delta \theta^2)^{-3}$).  The resulting profile of $\CE$ is again in remarkable agreement with that calculated in \cite{Athanasiou:2010pv}.

\section{Discussion}
\label{s:concl}

In the preceding section we have described the important features of the boundary energy density $\CE(t,r,\theta, \varphi)$ induced by a relativistic coiling string in the bulk.  We proposed approximating this by a superposition of gravitational shock waves produced by each string bit -- we dubbed this as `beaming mechanism' -- and calculated the consequences of this proposal.  In particular, we have obtained the characteristic spiral pattern of synchrotron radiation (cf.\ Figure \ref{f:bdy_GSW}).  We have calculated the spacing, phase, width, speed of propagation, as well as extent off the orbital plane, of each energy pulse, and found each of these characteristic features in complete quantitative agreement with the computations of \cite{Athanasiou:2010pv}.

Our results suggest that it is indeed valid to view the string's backreaction as given by superposition of GSWs.  Since each GSW is supported on a co-dimension 1 surface in AdS, this explains how the effect of a relativistic string deep in the bulk can remain so sharply localized all the way to the AdS boundary.  Consequently, this beaming mechanism elucidates the scale/radius duality, and indicates how to construct the counter-examples to the naive expectations mentioned in the Introduction.  From the CFT standpoint, this also bears on color transparency.

From a computational perspective, our proposal of estimating the backreaction by superposing GSWs presents a major computational simplification.  Instead of having to solve the full linearized Einstein's equations in the bulk, we merely need to find geodesics!  This provides a useful tool for understanding the radiation of a relativistic quark following an arbitrary trajectory.  In fact, generalizing the GSW construction to Schwarzschild-AdS background, one should likewise be able to estimate the quark's radiation in a thermal strongly-coupled plasma, such as studied in the context of a diffusion wake and sonic boom generated by a moving quark \cite{Friess:2006fk}.  Although here the string no longer remains relativistic everywhere, one could nevertheless use a series expansion in $1/\gamma$ to estimate the backreaction effects and their regime of validity.

Finally, from observational standpoint, it would be interesting to ask whether the beaming discussed above might lead to observational signatures from  relativistic cosmic strings, in addition to those produced by cusps and kinks  \cite{Damour:2001bk}.  We leave further discussion of these intriguing issues to \cite{Hubeny:2010xx}.

\subsection*{Acknowledgements}
\label{acks}

It is a pleasure to thank
Paul Chesler,
Roberto Emparan,
Ruth Gregory,
Gary Horowitz,
Hong Liu,
Shiraz Minwalla,
Rob Myers,
Joe Polchinski,
and Mukund Rangamani
for valuable discussions.
I would also like to thank MIT, UBC, Galileo Galilei Institute, and Imperial College London for their hospitality during this project.
This work is supported in part by a STFC Rolling Grant.

\providecommand{\href}[2]{#2}\begingroup\raggedright\endgroup

\end{document}